\documentclass[aps,pra,twocolumn,superscriptaddress,showpacs]{revtex4}

\usepackage{graphicx}
\usepackage{natbib}
\usepackage[breaklinks]{hyperref}
\usepackage{amsmath}
\usepackage{mathtools}
\usepackage[utf8]{inputenc} 

\newcommand{\hflev}[4]{\textit{#1}$_{#2/#3}$, \textit{F}=#4}
\begin{document}


\title{Polarization gradient cooling of single atoms in optical dipole traps}
\author{Yue-Sum Chin}
\affiliation{Centre for Quantum Technologies, 3 Science Drive 2, Singapore 117543}
\author{Matthias Steiner}
\affiliation{Centre for Quantum Technologies, 3 Science Drive 2, Singapore 117543}
\affiliation{Department of Physics, National University of Singapore, 2 Science Drive 3, Singapore 117542}
\author{Christian Kurtsiefer}
\affiliation{Centre for Quantum Technologies, 3 Science Drive 2, Singapore 117543}
\affiliation{Department of Physics, National University of Singapore, 2 Science Drive 3, Singapore 117542}
\email[]{christian.kurtsiefer@gmail.com}
\date{\today}

\begin{abstract}
We experimentally investigate  $\sigma^+$-$\sigma^-$ polarization gradient cooling~(PGC) of a single $^{87}$Rb atom in a tightly focused dipole trap and show that the cooling limit strongly depends on the polarization of the trapping field. 
For optimized cooling light power, the temperature of the atom reaches~$10.4(6)\,\mu$K in a linearly polarized trap, approximately five times lower than in a circularly polarized trap. 
The inhibition of PGC is qualitatively explained by the fictitious magnetic fields induced by the trapping field.  
We further demonstrate that switching the trap polarization from linear to circular after PGC induces only minor heating. 
\end{abstract}


\maketitle
Single neutral atoms in tightly focused optical traps are a promising platform for quantum information processing, quantum simulation, and to act as nodes in quantum networks~\cite{Kaufman2015,Barredo2015,Gaetan2009,Urban2009,Tey2008}. 
Many of these applications require the atom to be sufficiently cooled~\cite{Kaufman2012,Thompson2013,Sompet2017} in order to reduce the spatial spread~\cite{Chin2017}, increase the coherence time~\cite{Yavuz2006,Rosenfeld2008}, or use quantum mechanical properties of the atomic motion~\cite{Kaufman2014}.
Optically confined atoms, like free atoms, can be cooled to sub-Doppler temperatures by polarization gradient cooling~(PGC)~\cite{Weiss1989,Ungar1989,Dalibard1989}. 
However,  despite its practical relevance, the influence of the optical trap on the efficiency of PGC is relatively unexplored; 
for example, reported temperatures for the commonly used atomic species $^{87}$Rb vary by an order of magnitude for similar experimental configurations~\cite{Rosenfeld2008,Tuchendler2008,Kaufman2012}.  
In this work, we experimentally address this topic and investigate PGC of single atoms in a mK-deep far off-resonant optical dipole trap~(FORT).  
In particular, we consider the configuration of counter-propagating beams of opposite circular polarizations, referred to as $\sigma^+$-$\sigma^-$ PGC, and explore the dependency of the cooling limit on the polarization of the trapping field. 

Shortly after the initial demonstrations of $\sigma^+$-$\sigma^-$ PGC, it became clear that, while this cooling technique is in general robust against small variations of the experimental parameters, it is very sensitive to magnetic fields~\cite{Shang1990,Shang1991,Walhout1992,Walhout1996,Werner1992,Chang2002}. 
The reason for the detrimental effect of magnetic fields is that $\sigma^+$-$\sigma^-$ PGC is based on velocity-selective Raman transitions, which redistribute population within the spin states of the ground state manifold. 
The associated Zeeman effect shifts the Raman resonance, and thus the atoms are no longer cooled toward zero velocity but to a finite velocity at which the Doppler shift compensates the Zeeman shift.  
\begin{figure} 
\centering
  \includegraphics[width=\columnwidth]{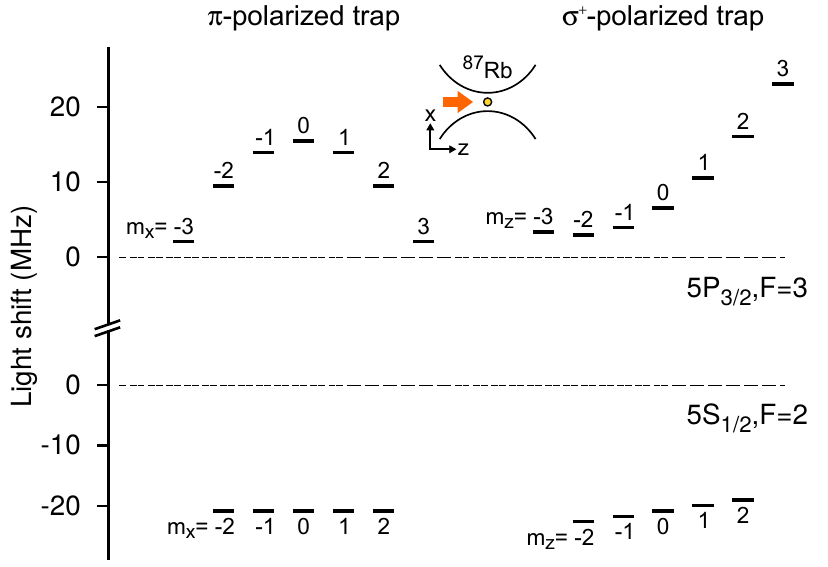}
  \caption{\label{fig:levelscheme}Energy level scheme for the \mbox{5\hflev{S}{1}{2}{2}} to  \mbox{5\hflev{P}{3}{2}{3}} transition near 780\,nm of a $^{87}$Rb atom in a $\pi\textrm{-polarized}$ (parallel to x axis) and a $\sigma^+\textrm{-polarized}$ FORT.
  Inset illustrates the geometrical arrangement: The trapping beam propagates along the z axis. 
   }
\end{figure}

Similarly, the energy levels of the cooling transition are shifted for an atom in a FORT. 
In our experiment $\sigma^+$-$\sigma^-$ PGC of $^{87}$Rb atoms is performed on the closed \mbox{5\hflev{S}{1}{2}{2}} to \mbox{5\hflev{P}{3}{2}{3}} transition near 780\,nm.
Figure~\ref{fig:levelscheme} shows the calculated light shifts for a linearly~$\pi$-polarized and circularly~$\sigma^+$-polarized FORT operating at 851\,nm with a trap depth of $U_0=k_B\times 1$\,mK~\cite{Safronova2011,Shih2013}. 
In a $\pi$-polarized trap, the energy shift is the same for all spin states within the ground state \mbox{5\hflev{S}{1}{2}{2}} manifold~\cite{Deutsch1998,Corwin1999}. 
This degeneracy is lifted in a $\sigma^+$-polarized trap, where the trapping field acts as a `fictitious magnetic field' pointing in the direction of propagation~\cite{Cohen-Tannoudji1972}.  
Both $\pi$ and $\sigma^+$-polarized light lifts the degeneracy of the Zeeman manifold in the excited state~\mbox{5\hflev{P}{3}{2}{3}}.

\begin{figure} 
\centering
  \includegraphics[width=\columnwidth]{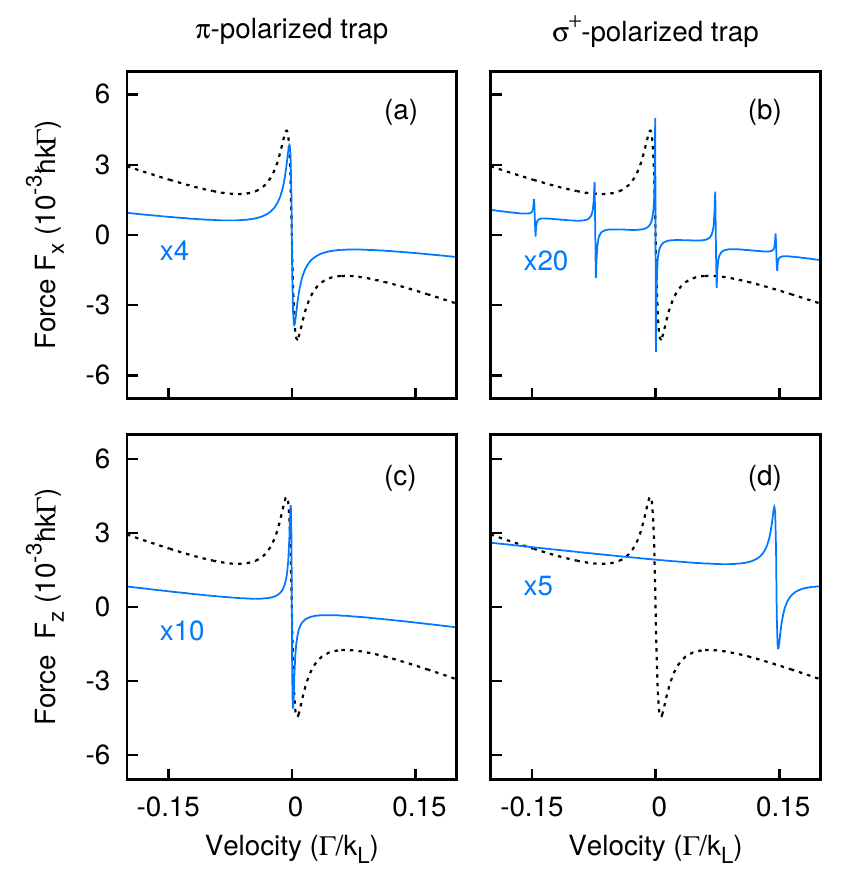}
  \caption{\label{fig:pgc_force}
  Calculated force on an atom of fixed velocity moving through a $\sigma^+$-$\sigma^-$ PGC field for different axes and FORT polarizations. 
    Both beams of the PGC field have a Rabi frequency $\Omega=\Gamma_0/2$ and are red-detuned from the natural transition frequency by $\Delta=-3\Gamma_0$, where~$\Gamma_0=2\pi \times 6.07\,$MHz is the natural linewidth. 
    Black dashed and blue solid lines indicate the force for a free and a trapped atom, respectively.
  (a)~$\pi$-polarized trap, PGC field along x axis. (b)~$\sigma^+$-polarized trap, PGC field along x axis. 
  (c)~$\pi$-polarized trap, PGC field along z axis. (d)~$\sigma^+$-polarized trap, PGC field along z axis. 
      }
\end{figure}
To qualitatively understand the effect of the light shifts on PGC, we calculate the force an atom of fixed velocity experiences when traveling across a $\sigma^+$-$\sigma^-$ PGC field in the FORT. 
We use a semi-classical description which defines the force~$F$ on an atom as
the expectation value of the quantum mechanical force operator, $F=-\langle \nabla \hat{H} \rangle$~\cite{Metcalf2001}. 
The total Hamiltonian~$\hat{H}=\hat{H}_0+\hat{H}_\textrm{int}$ consists of two parts: 
(1)~a spatially independent Hamiltonian~$\hat{H_0}$ which contains the energy levels of the cooling transition including the light shifts induced by the trap and (2)~a Hamiltonian which describes the interaction with the near-resonant
PGC field, $\hat{H}_\textrm{int}=-\frac{\hbar}{2} \left( \Omega_+(\vec{r}) \hat{A}_+ + \Omega_-(\vec{r}) \hat{A}_- + \Omega_\pi(\vec{r}) \hat{A}_\pi \right)+h.c.$, where $\Omega_+$, $\Omega_-$ and $\Omega_\pi$ are the spatially dependent Rabi frequencies for $\sigma^+$, $\sigma^-$ and $\pi$-polarized light, with $\hat{A}_+$, $\hat{A}_-$ and $\hat{A}_\pi$ as the atomic lowering operators for the respective polarizations. 
For a given atomic velocity, we solve the corresponding master equation, $\dot{\rho}=-\frac{i}{\hbar}[\rho,\hat{H}] + \mathcal{L}[\rho]$ by the matrix continued fraction method ($\mathcal{L}[\rho]$ is the Lindblad superoperator accounting for spontaneous emission)~\cite{Minogin1979,Tan1999,Lee2013}. 
We then compute the steady-state force averaged over the travel through one cycle of the light. 

For a free atom, the simulation shows a steep slope of the force around zero velocity, which is a hallmark of sub-Doppler cooling (Fig.~\ref{fig:pgc_force}, black solid line). 
For an atom confined in a FORT, the force depends strongly on the trap polarization and the angle between the trapping beam and the PGC field. 
Figure~\ref{fig:pgc_force} shows the force for two polarizations, linear~$\pi$ along the x axis and circular~$\sigma^+$, as well as two orientations for the PGC field, parallel and perpendicular to the trapping beam. 
In the $\pi$-polarized trap~[Fig.~\ref{fig:pgc_force}(a) and~(c)], the persisting steep slope of the force around zero velocity indicates that the PGC is little affected by the trap, aside from a narrowing of the sub-Doppler feature due to the increased detuning from the cooling transition.  
The $\sigma^+$-polarized trap exhibits five resonances when the PGC field is perpendicular to the trapping beam~[Fig.~\ref{fig:pgc_force}(b)]. 
These velocity selective resonances correspond to Raman transitions between ground state sublevels, known from PGC cooling in strong transverse magnetic fields~\cite{Walhout1996}. 
For a PGC field parallel to the trapping beam, only one Raman transition can
be brought into resonance by the motion of the
atom~[Fig.~\ref{fig:pgc_force}(d)] --- a situation which resembles PGC in longitudinal magnetic fields~\cite{Walhout1992}. 
Although this simple 1-D model of the force cannot predict the final temperatures in the actual experiment, it indicates that PGC works in $\pi$-polarized traps, but is strongly compromised in mK-deep $\sigma^+$-polarized traps. 
\begin{figure} 
\centering
  \includegraphics[width=\columnwidth]{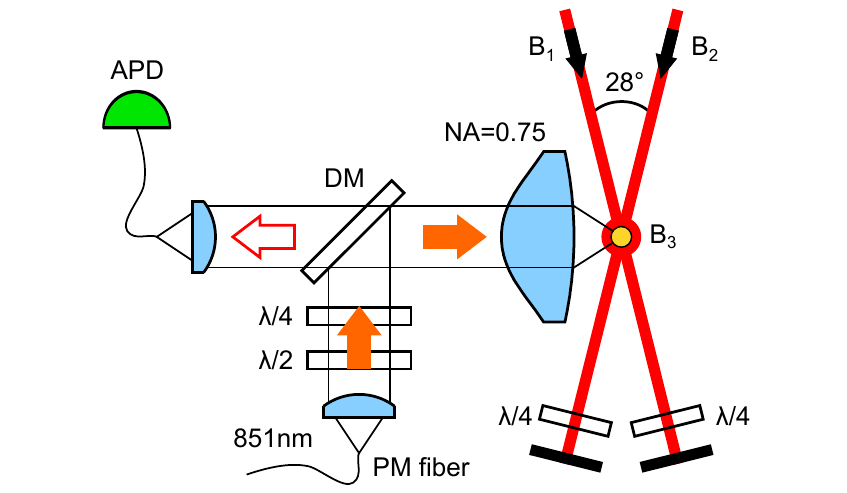}
  \caption{\label{fig:setup}
  Optical setup for trapping, polarization gradient cooling, and fluorescence detection of a single atom. 
  APD:~avalanche photodetector, DM:~dichroic mirror, $\lambda/4$:~quarter-wave plate, $\lambda/2$:~half-wave plate, B:~beam consisting of 780\,nm cooling light and 795\,nm repumping light with a waist of 1\,mm. 
  B$_3$ is perpendicular to B$_1$ and B$_2$. 
  }
\end{figure}

Our experiment starts with a magneto-optical trap (MOT) from which we load a single $^{87}$Rb atom into a red detuned FORT by light-induced collisions~\cite{Schlosser2001,Schlosser2002}. 
The dipole trap is formed by 851\,nm light that is tightly focused by a high numerical aperture lens (NA=0.75, focal length~$f$=5.95\,mm, see Fig.~\ref{fig:setup}), resulting in a trap depth of $U_0=k_\textrm{B}\times1.88(1)$\,mK, with radial frequencies  $\omega_{r}/2\pi=113(1)$\,kHz, $\omega_{r'}/2\pi=98(1)$\,kHz, and an axial frequency $\omega_z/2\pi=12.6(1)$\,kHz~\cite{Chin2017,Chin2017a}. 
Part of the atomic fluorescence is collected by the same lens and coupled to a single mode fiber connected to an avalanche photodetector. 
We use the same light for the MOT and PGC, provided by three circularly polarized beams, which are retroreflected with opposite polarization. 
Two of these beams~B$_1$,B$_2$ are non-orthogonal, and have a propagation component along the direction of the trapping beam to ensure cooling along that axis. 
The third beam~B$_3$ is orthogonal to these two beams and carries twice as
much power.
We modulate the mirror position of the cooling beams with an amplitude of $1\,\mu$m at 100\,Hz to average the interference pattern of the cooling light over the atom position~\cite{Kaufman2012}. 
The frequency of the cooling light is red-detuned from the natural transition frequency by typically $\Delta=-3\Gamma_0$. 
In addition, all beams carry repumping light nearly resonant with the D$_1$ line at 795\,nm to clear out the \mbox{5\hflev{S}{1}{2}{1}} population. 

Once an atom is trapped, we turn off the magnetic quadrupole field and apply PGC for 10\,ms. 
Subsequently, we use a `release and recapture' method to measure the temperature of the atoms~\cite{Tuchendler2008}: 
The cooling and repumping light is switched off and the atom is released from the trap for an interval~$t_\textrm{r}$ by interrupting the trapping beam.  
We detect the atomic fluorescence by switching back on cooling and repumping light to determine whether the atom was recaptured. 
For a set of 11 different release intervals~$t_\textrm{r}$, we repeat each experiment several hundred times to obtain an estimate of the recapture probability. 
Finally, we extract the temperature by comparing the recapture probabilities to a Monte-Carlo simulation~\cite{Tuchendler2008}. 

\begin{figure}[h]
\centering
  \includegraphics[width=1\columnwidth]{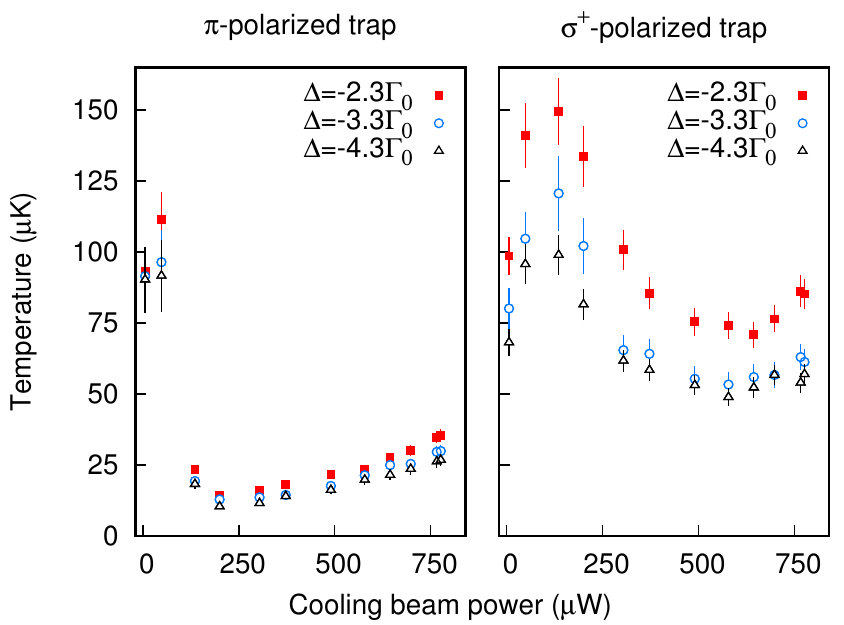}
  \caption{\label{fig:pgc_intensity}
  Temperature of the atoms after PGC over the total cooling beam power in B$_1$, B$_2$, and B$_3$.
  Error bars represent one standard deviation.
  }
\end{figure}

\begin{figure} 
\centering
  \includegraphics[width=\columnwidth]{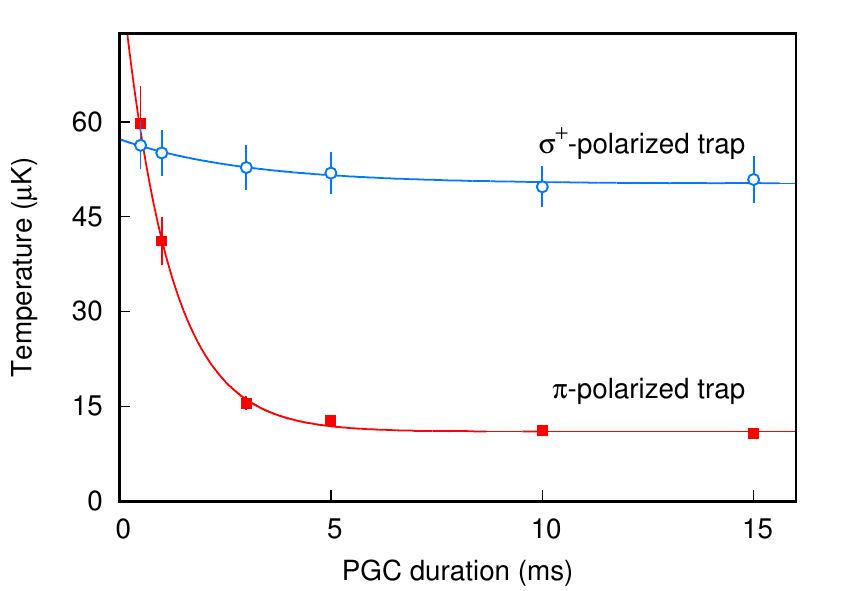}
  \caption{\label{fig:pgc_duration}
  Temperature of the atoms after PGC for a varying cooling duration. 
  Optimal cooling beam power is used respectively for both the $\pi$-polarized trap (red square) and the $\sigma^+$-polarized trap (blue circle). 
  Solid lines are fits to exponentials. 
  Error bars represent one standard deviation. 
  }
\end{figure}

We compare PGC in a $\pi$-polarized (parallel to beam~B$_3$) trap with that in a $\sigma^+$-polarized trap. 
To optimize the cooling parameters to reach the lowest temperatures, we adjust the cooling beam power and frequency~(Fig.~\ref{fig:pgc_intensity}).
We observe the typical PGC behavior of lower temperatures for larger detunings of the cooling beam and an optimal cooling power below which the temperature increases sharply~\cite{Salomon1990,Gerz1993}. 
This behavior is more pronounced in the $\pi$-polarized trap than in the $\sigma^+$-polarized trap. 
The lowest temperature is achieved in the $\pi$-polarized trap at~$10.4(6)\,\mu$K, which is approximately 5 times lower than the lowest temperature observed in the $\sigma^+$-polarized trap at~$49(3)\,\mu$K. 
Figure~\ref{fig:pgc_duration} shows the temperature of the atoms after a
variable time of PGC, measured with the respective optimal cooling beam power. 
In the $\pi$-polarized trap, the atom is quickly ($1/e$-time constant of $1.1(1)$\,ms) cooled to low temperatures, whereas in the $\sigma^+$-polarized trap PGC is inhibited and the atom remains close to the initial temperature. 

\begin{figure} 
\centering
  \includegraphics[width=\columnwidth]{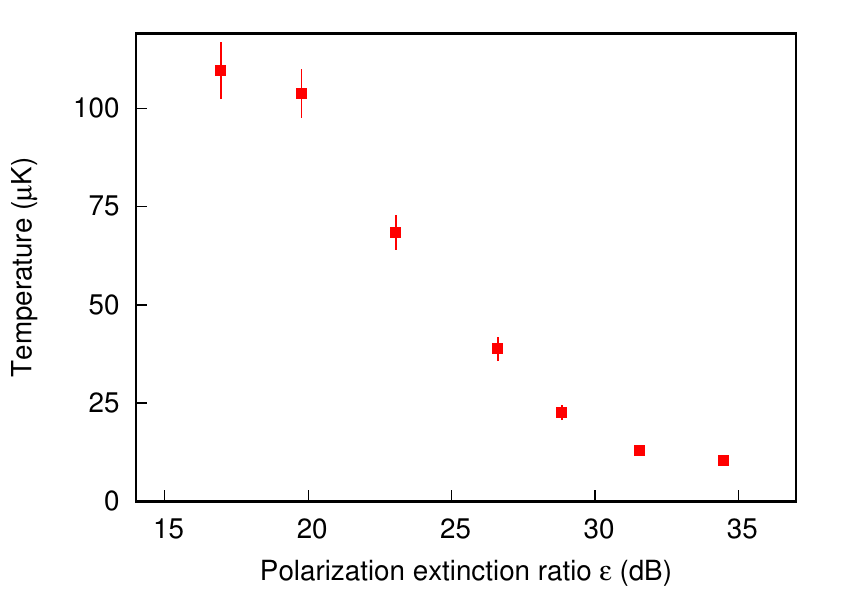}
  \caption{\label{fig:ellipticity}
  Temperature of the atoms after PGC in a $\pi$-polarized trap depending on the polarization extinction ratio. 
  The cooling beam power is optimized for the highest value of~$\epsilon$. 
  Error bars represent one standard deviation. 
}
\end{figure}

To test how sensitive the cooling in the $\pi$-polarized trap is to imperfections of the polarization, we deliberately introduce a slight ellipticity. 
The quality of the polarization here is quantified as the polarization
extinction ratio $\epsilon=10{\textrm{dB}}\,\textrm{log}_{10}(P_\textrm{max}/P_\textrm{min})$,
where $P_\textrm{max}$ and $P_\textrm{min}$ are the maximum and minimum transmitted power through a rotating film-polarizer. 
As shown in Fig.~\ref{fig:ellipticity}, we find a high sensitivity of the PGC to the purity of the linear polarization. 
Already at $\epsilon=32$\,dB, the temperature $13(1)\,\mu$K is notably higher compared to $10.4(6)\,\mu$K at $\epsilon=35$\,dB. 
We do not expect much lower temperatures for polarization extinction ratios above $\epsilon=35$\,dB because for our lowest observed temperature of $10.4(6)\,\mu$K, the mean phonon number of the radial mode~$\bar{n}_{r}= (e^{\hbar\omega_{r}/k_B T}-1)^{-1}=1.5(1)$ is close to the theoretical limit of $\bar{n}\approx1$~\cite{Wineland1992,Cirac1993}.
Recently, a similar value for the mean phonon number has also been observed for PGC of trapped ions~\cite{Ejtemaee2016}.

Finally, we demonstrate that switching the trap polarization from linear to circular after PGC induces only minor heating. 
The polarization switch is implemented with a free-space transverse
  electro-optical polarization modulator. 
Insertion of the polarization modulator and additional waveplates compromises
the purity of the $\pi$-polarization, leading to a polarization extinction ratio~$\epsilon=33$\,dB. 
Consequently, we find a slightly increased temperature of $13.1(9)\,\mu$K after PGC cooling in the $\pi$-polarized trap. 
Next, we switch the polarization after PGC and perform the release--recapture experiment in the $\sigma^+$-polarized trap. 
We observe a marginally increased temperature to $13.8(7)\,\mu$K, which is likely caused by an approximately 1\% change in dipole trap power after the switching.
Nevertheless, the achieved temperature is a significant improvement over PGC in a $\sigma^+$-polarized trap. 

In summary, we demonstrated that  $\sigma^+$-$\sigma^-$~polarization gradient cooling in a linearly polarized dipole trap leads to a lower atom temperature  compared to a circularly polarized trap. 
The cooling limit shows a strong sensitivity on the purity of the linear polarization; 
we measure a temperature increase from $10.4(6)\,\mu$K to $13(1)\,\mu$K when we reduce the polarization extinction ratio from  $35$\,dB to $32$\,dB. 
In this sense our results agree with the review article~\cite{Grimm2000}, published almost two decades ago, stating `\dots linearly polarized light is usually the right choice for a dipole trap\dots '. 
However, in practice the choice of the trap polarization is often set for other reasons than to optimize the PGC. 
For example, in experiments testing the interaction of atoms with tightly focused light employ co-propagating FORT and probe light, a circularly polarized trap is necessary to efficiently drive the strong cycling transition~\cite{Tey2008}. 
Such experiments can benefit from dynamical control of the trap polarization, i.e., performing PGC in a linearly polarized trap before conducting the experiment in a circularly polarized trap~\cite{Chin2017}. 

\begin{acknowledgments}
We acknowledge the support of this work by the Ministry of Education in
Singapore (AcRF Tier 1) and the National Research Foundation, Prime Minister's
office.
M.\,Steiner acknowledges support by the Lee Kuan Yew Postdoctoral Fellowship.
\end{acknowledgments}

\bibliographystyle{apsrev4-1}
%

\end{document}